\documentclass[aps,showpacs,twocolumn]{revtex4}
\usepackage{epsfig}
\usepackage{color}
\usepackage{amsmath}
\usepackage{amssymb}
\usepackage{bbm}

\newcommand{\be}{\begin{equation}}
\newcommand{\ee}{\end{equation}}
\newcommand{\bea}{\begin{eqnarray}}
\newcommand{\beaa}{\begin{eqnarray*}}
\newcommand{\eea}{\end{eqnarray}}
\newcommand{\eeaa}{\end{eqnarray*}}

\begin{document}

\title{\bf\Large {Coexistence of superconductivity and magnetism in spin-fermion model of ferrimagnetic spinel in an external magnetic field}}

\author{ Naoum Karchev\cite{byline}}

\affiliation{Department of Physics, University of Sofia, 1164 Sofia, Bulgaria}

\begin{abstract}
A two-sublattice spin-fermion model of ferrimagnetic spinel, with spin-$1/2$ itinerant electrons at the sublattice $A$ site and spin-$s$ localized electrons  at the sublattice $B$ site
is considered. The exchange between itinerant and localized electrons is antiferromanetic. As a result the external magnetic field, applied along the magnetization of the localized electrons, compensates the Zeeman splitting due to the spin-fermion exchange and magnon-fermion interaction induces spin anti-parallel p-wave superconductivity which coexists with magnetism.
We have obtained five characteristic values of the applied field (in units of energy) $H_{cr1}<H_3<H_0<H_4<H_{cr2}$.
At $H_0$ the external magnetic field compensates the Zeeman splitting. When $H_{cr1}<H<H_{cr2}$ the spin antiparallel p-wave superconductivity with $T_{1u}$ configuration coexists  with magnetism. The superconductor to normal magnet transition at finite temperature is second order when $H$ runs the interval $(H_3,H_4)$. It is an abrupt transition when $H_{cr1}<H<H_3$ or $H_4<H<H_{cr2}$. This is proved calculating the temperature dependence of the gap for three different values of the external magnetic field $H_{cr1}<H<H_3$, $H_4<H<H_{cr2}$ and $H=H_0$. In the first two cases the abrupt fall to zero of the gap at superconducting critical temperature shows that the superconductor to normal magnet transition is first order. The Hubbard term (Coulomb repulsion), in a weak coupling regime, does not affect significantly the magnon induced superconductivity. Relying on the above results one can formulate a recipe for preparing a superconductor from ferrimagnetic spinel: i) hydrostatic pressure above the critical value of insulator-metal transition. ii) external magnetic field along the sublattice magnetization with higher amplitude.

\end{abstract}

\pacs{75.50.Gg,74.20.Mn,74.20.Rp}

\maketitle

Electron-phonon mechanism of superconductivity \cite{Bardeen50,Frohlich50,Cooper56,BCS57} has been developed  to explain the pairing in a large variety of materials, from $Hg$ and $Al$ to recently discovered $MgB_2$ \cite{MgB2}. The discovery of superconductivity in $LaBaCuO$ \cite{Bednorz86} and in other cuprates, as well as discovery of superconductivity in $Fe$-based pnictides \cite{Fe} established another direction of research in this field. The boson-fermion models still remained respected but bosons are not lattice vibrations.

Alternatively, the possibility of an electronic pairing mechanism in systems with rotational invariance was put forward in a seminal paper by Kohn and Luttinger \cite{KL65,Luttinger66,FL68}. Although the  bare interaction among electrons is repulsive, there is an effective attractive interaction that arise at higher order of perturbation theory. The Kohn-Luttinger instability of a three-dimensional rotationally invariant system results in the formation of a unconventional superconducting ground state due to the peak in the particle-hole susceptibility near zero wave vector. The works \cite{Chubukov92,ZS96,HSR02,RKS10,AK11} have made significant progress in our understanding of superconductivity from repulsive interaction.

There are many experiments addressing external magnetic field induced, enhanced or reentered superconductivity. Experimentally, an anomalous enhancement of $H_{c2}(T )$ was first reported by Fischer et al\cite{Fischer75}. An increase of about $50-100\, kG$ of the upper critical field is observed in $Sn_{1.2(1-x)}Eu_{x}Mo_{6.35}S_8$ and $Pb_{1-x}Eu_x Mo_{6.35}S_8$ with respect to the compounds without Eu. The overall feature of the field-induced superconducting phase is well understood by theory based on the Jaccarino- Peter (JP) compensation mechanism\cite{JP62}.

In a rare earth ferromagnetic metal the conduction electrons are in an effective field due to the exchange interaction with the rare earth spins. It is in general so large as to inhibit the occurrence of superconductivity. For some systems the exchange interaction have a negative sign. This  allows for the conduction electron polarization to be canceled by an external magnetic field so that if, in addition these metals possess phonon-induced
attractive electron-electron interaction, superconductivity  occurs in the compensation region. If the effective field is not large the coexistence of superconductivity end magnetic order is possible and the external magnetic field enhances the superconductivity. The effect can also be observed in a paramagnet since the strong external field will in any case polarize the localized magnetic moments at low temperature, and thus
produce the necessary ferromagnetic alignment \cite{Fischer72}.Therefore, superconductivity can occur in two domains: one at the low field, where the pair-breaking field is still small, and one at the high field in the compensation region. The field reentrance of superconductivity was first reported in \cite{Wolf82,Meul84}.

The JP compensation mechanism was originally proposed to explain the superconductivity in some pseudoternary materials. Recently, the JP effect has been proven to be responsible for the magnetic-field-induced superconductivity in the organic superconductor $\lambda-(BETS)_2FeCl_4$ \cite{Uji01,Balicas01}.

The superconductivity in the Jaccarino- Peter theory is induced by phonon fluctuations and spin fluctuations (magnons) weaken the spin singlet Cooper pairing. In the present paper we consider magnon induced superconductivity based on the compensation mechanism\cite{JP62}.
We study the conditions for the coexistence of superconductivity and magnetism in a spin-fermion system which is a prototype model of itinerant ferrimagnetic spinel. A two-sublattice system is defined on a body centered cubic  lattice, with spin-$1/2$ itinerant electrons at the sublattice $A$ site and spin-$s$ localized electrons at the sublattice $B$ site. The subtle point is the exchange between itinerant and localized electrons which is antiferromanetic and applying an external magnetic field along the magnetization of the localized electrons one can compensate the Zeeman splitting due to the spin-fermion exchange. Then, magnon-fermion interaction induces spin anti-parallel p-wave superconductivity, with $T_{1u}$ configuration, which coexists with magnetism. We have studied the superconducting gap as a function of applied magnetic field and temperature.
The Coulomb repulsion, in a weak coupling regime, does not affect significantly the magnon induced superconductivity.

Relying on the above results one can formulate a recipe for preparing a superconductor from ferrimagnetic spinel: i) hydrostatic pressure above the critical value of insulator-metal transition. ii) external magnetic field along the sublattice magnetization with higher amplitude.
In favor of this recipe one can mention that metallization in magnetite $Fe_3O_4$  is found  under a pressure above $8 GPa$ \cite{Fe3O401,Fe3O402,Fe3O408}. While the model under consideration does not match well the $Fe_3O_4$ system we expect to find superconductivity  applying external magnetic field along sublattice B magnetization, when the hydrostatic pressure is above the critical one.

On the other hand, there are spinel compounds well known as superconductors at ambient pressure $CuRh_2Se_4$,$CuRh_2S_4$\cite{spinelSC1,spinelSC2,spinelSC3,spinelSC4,spinelSC5,spinelSC6}. The results of the present paper inspire that applying external magnetic field one can expect an enhancement of the superconducting transition temperature $T_{sc}$. This is quite specific phenomenon for the spinel superconductivity and it deserves to be experimentally verified.

The Hamiltonian of the spin-fermion model of ferrimagnetic spinel defined on a body centered cubic lattice is
\bea \label{IFerri1}\nonumber
h  = & - & t\sum\limits_{\ll ij \gg _A } {\left( {c_{i\sigma }^ + c_{j\sigma } + h.c.} \right)}
  -\mu \sum\limits_{i\in A} {n_i
} \\ & - & J^B\sum\limits_{  \ll  ij  \gg_B  } {{\bf S_i^B}
\cdot {\bf S_j^B}}
  + J\sum\limits_{  \langle  ij  \rangle } {{\bf
S_i^A}}\cdot {\bf S_j^B}  \\
&-& H \sum\limits_{i\in A} {S^{zA}_{i}}-H \sum\limits_{i\in B} {S^{zB}_i}, \nonumber
\eea
where $S^{\nu A}_i=\frac 12\sum\limits_{\sigma\sigma'}c^+_{i\sigma}\tau^{\nu}_{\sigma\sigma'}c^{\phantom +}_{i\sigma'}$, with the Pauli
matrices $(\tau^x,\tau^y,\tau^z)$, is the spin of the itinerant
electrons at the sublattice $A$ site , ${\bf S_i^B}$ is the spin of the localized electrons  at the sublattice $B$ site, $\mu$
is the chemical potential, and $n_i=c^+_{i\sigma}c_{i\sigma}$. The
sums are over all sites of a body centered cubic lattice, $\langle i,j\rangle$ denotes the sum over the nearest neighbors, while $ \ll  ij  \gg_A$ and  $\ll  ij  \gg_B$ are sums over all sites of sublattice $A$ and $B$ respectively. The Heisenberg term $(J^B > 0)$ describes ferromagnetic Heisenberg
exchange between localized electrons and $J>0$ is the antiferromagnetic exchange constant between localized and itinerant electrons. $H>0$ is the Zeeman splitting energy due to the external magnetic field (magnetic field in units of energy).

To proceed we use the Holstein-Primakoff representation of the spin operators of localized electrons ${\bf S^B_j}(a^+_j,a_j)$, where $a^+_j,\,a_j$
are Bose fields. In terms of these fields and keeping only the quadratic terms, the  Hamiltonian
Eq.(\ref{IFerri1}) is a sum of three terms
\be\label{IFerri2} h=h_b+h_f+h_{bf},\ee
where
\bea\label{IFerri3} \nonumber
  h_b & = &  s\,J^B\sum\limits_{  \ll  ij  \gg_B
 }(a_i^+a_i+a_j^+a_j-a_j^+a_i-a_i^+a_j)\\
 & & + H \sum\limits_{i\in B} {a_i^+a_i} \nonumber \\
 h_f & = & -t\sum\limits_{\ll ij \gg _A } {\left( {c_{i\sigma }^ + c_{j\sigma } + h.c.} \right)}
  -\mu \sum\limits_{i\in A} {n_i} \\
 & & +  (4Js-H)\sum\limits_{i\in A} \frac 12 {c^+_{i\sigma}\tau^{3}_{\sigma\sigma'}c^{\phantom +}_{i\sigma'}} \nonumber \\
 h_{bf} & = & \sqrt{\frac{s}{2}}J\sum\limits_{  \langle  ij  \rangle }\left(c_{i\downarrow }^ + c_{i\uparrow }a_j+c_{i\uparrow}^ + c_{i\downarrow }a_j^+\right)\nonumber \eea
 In momentum space representation, the Hamiltonian reads
 \bea\label{IFerri4} \nonumber
h_b & = &  \sum\limits_{k\in B_r} \varepsilon_k a_k^+ a_k \\
h_f & = &  \sum\limits_{k\in B_r \sigma} \varepsilon_{k \sigma} c_{k \sigma}^+ c_{k \sigma} \\
h_{bf} & = & \frac {4J\sqrt{2s}}{\sqrt{N}}\sum\limits_{k q p \in B_r } \delta (p-q-k)\cos\frac {k_x}{2} \cos\frac {k_y}{2} \cos\frac {k_z}{2} \nonumber \\ & \times & \left(c_{p\downarrow }^ + c_{q\uparrow }a_k+c_{q\uparrow}^ + c_{p\downarrow }a_k^+\right) , \nonumber \eea
with bose $\varepsilon_k$ and fermi $\varepsilon_{k \sigma}$ dispersions
\bea\label{IFerri5}
\varepsilon_k & = & 2sJ^B\left ( 3-\cos k_x-\cos k_y-\cos k_z \right)+H \\
\varepsilon_{k \uparrow} & = & -2t\left ( \cos k_x+\cos k_y+\cos k_z \right)-\mu+\frac {4sJ-H}{2} \nonumber \\
\varepsilon_{k \downarrow} & = & -2t\left ( \cos k_x+\cos k_y+\cos k_z \right)-\mu-\frac {4sJ-H}{2} \nonumber \eea
The two equivalent sublattices A and B of the body center cubic lattice are simple cubic lattices. Therefor the wave vectors $p,q,k$ run over the first Brillouin zone of a cubic lattice $B_r$ .

Let us average in the subspace of Bosons $(a^+,a)$ ( to integrate the Bosons in the path integral approach). In static approximation one obtains an effective fermion theory with Hamiltonian $h_{eff}=h_f+h_{int}$, where $h_f$ is the free fermion Hamiltonian Eq.\eqref{IFerri4} and the magnon-induced four-fermion interaction is
\bea\label{IFerri6}
h_{int} = & - & \frac 1N \sum\limits_{k_i p_i \in B_r} \delta (k_1-k_2-p_1+p_2) \nonumber \\
 & \times & V_{k_1-k_2} c_{k_1\downarrow}^+c_{k_2\uparrow}c_{p_2\uparrow}^+c_{p_1\downarrow} \eea
with potential
\be\label{IFerri7}
V_k = \frac {32 s J^2\left (\cos\frac {k_x}{2} \cos\frac {k_y}{2} \cos\frac {k_z}{2}\right)^2}{ 2sJ^B\left ( 3-\cos k_x-\cos k_y-\cos k_z \right)+H} \ee

Following standard procedure one obtains the effective Hamiltonian in the Hartree-Fock approximation
\be\label{IFerri8}
h_{HF}=  \sum\limits_{k\in B_r}\left[ \varepsilon_{k \sigma} c_{k \sigma}^+ c_{k \sigma}+\Delta_k c_{-k\downarrow}^+c_{k\uparrow}+\Delta_k^+c_{k\uparrow}c_{-k\downarrow}\right], \ee
with gap function
\be\label{IFerri9}
\Delta_k=\frac 1N \sum\limits_{p\in B_r}<c_{-p\uparrow}c_{p\downarrow}> V_{p-k} \ee
The Hamiltonian can be written in a diagonal form by means of Bogoliubov excitations $\alpha^+,\alpha,\beta^+,\beta$, which have the following dispersions:
\bea\label{IFerri10}
E^{\alpha}_k & = & \frac 12 \left[\varepsilon_{k\uparrow}-\varepsilon_{k\downarrow}+\sqrt{(\varepsilon_{k\uparrow}+\varepsilon_{k\downarrow})^2+4|\Delta_k|^2}\right] \\
E^{\beta}_k & = & \frac 12 \left[-\varepsilon_{k\uparrow}+\varepsilon_{k\downarrow}+\sqrt{(\varepsilon_{k\uparrow}+\varepsilon_{k\downarrow})^2+4|\Delta_k|^2}\right].\nonumber \eea
In terms of the new excitations the gap equation reads
\bea\label{IFerri11}\nonumber
\Delta_k= & - & \frac 1N \sum\limits_{p\in B_r}V_{k+p}\frac {\Delta_p}{\sqrt{(\varepsilon_{p\uparrow}+\varepsilon_{p\downarrow})^2+4|\Delta_p|^2}} \\
& \times & \left(1-<\alpha^+_p\alpha_p>-<\beta^+_p\beta_p>\right), \eea
where $<\alpha^+_p\alpha_p>$ and $<\beta^+_p\beta_p>$ are fermi functions for Bogoliubov fermions.

Straightforward calculations show that equation \eqref{IFerri11} has not spin-singlet $\Delta_{-k}=\Delta_{k}$ solutions. Having in mind that sublattices are simple cubic lattices and following the classifications for spin-triplet gap functions $\Delta_{-k}=-\Delta_{k}$, we obtained that $A_{1u}$ state $\Delta_k=\Delta \sin k_x\sin k_y\sin k_z$ is not solution of the equation \eqref{IFerri11} too. The gap function with $T_{1u}$ configuration
 \be\label{IFerri12} \Delta_k=\Delta\left(\sin k_x+\sin k_y+\sin k_z) \right) \ee
is a solution of the gap equation for some values of the external magnetic field and temperature.

\begin{figure}[!t]
\epsfxsize=\linewidth
\epsfbox{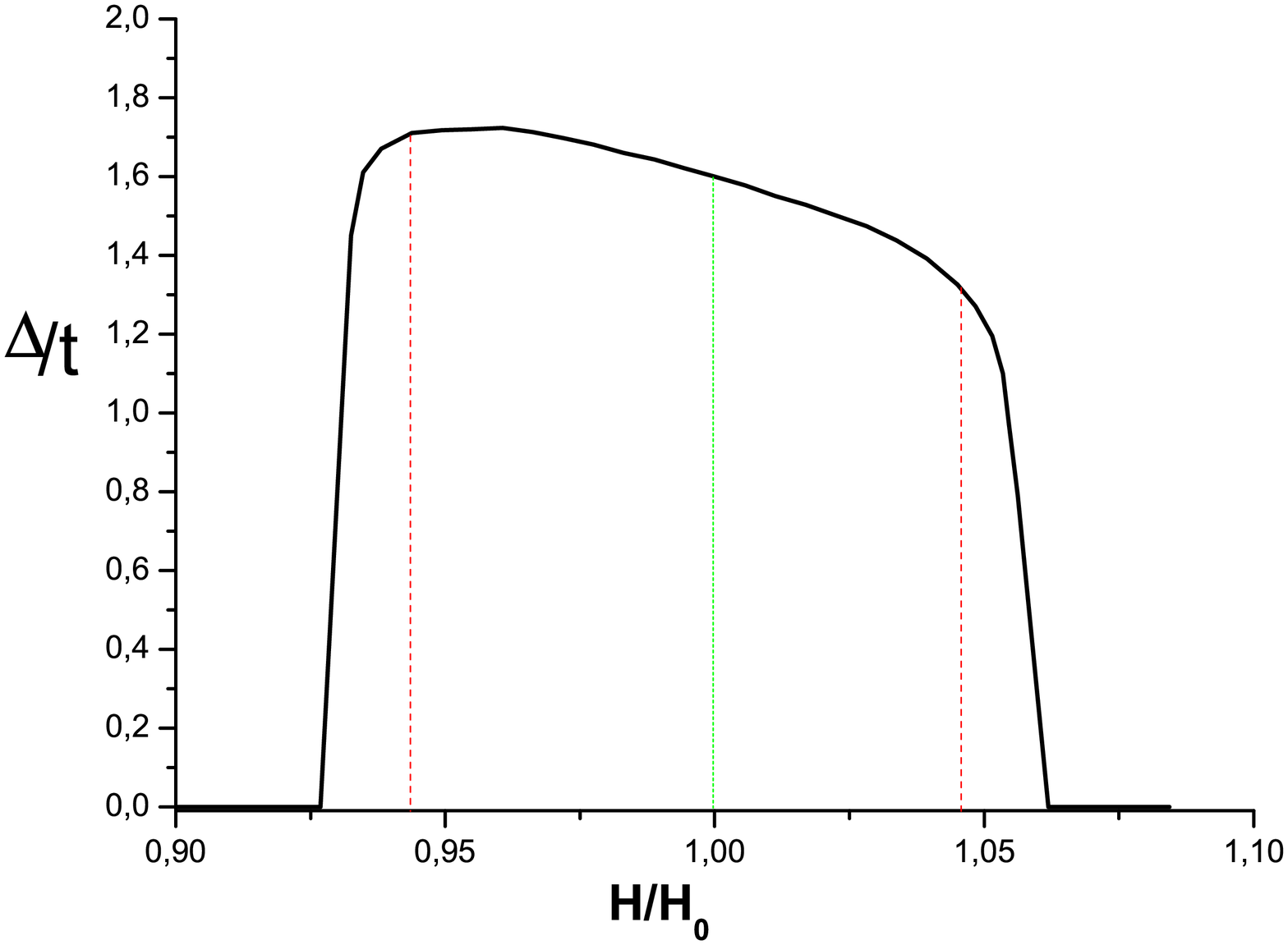} \caption{(Color online)\,\,Dimensionless gap $\Delta/t$ as a function of $H/H_0$. The critical values are $H_{cr1}/H_0=0.927$ and $H_{cr2}/H_0=1.062$. The red lines correspond to $H_3/H_0=0.944$ and $H_4/H_0=1.045$. When $H_3/H_0<H/H_0<H_4/H_0$ the thermal superconductor to normal magnet transition is second order. In other cases it is abrupt.   }\label{gap-h}
\end{figure}

It follows from equations \eqref{IFerri5}, that the external magnetic field (in units of energy) compensates the Zeeman splitting, due to the spin-fermion exchange, at $H=H_0=4sJ$. We calculate the gap parameter $\Delta$, from Eq.\eqref{IFerri11}, as a function of $H/H_0$, setting the chemical potential $\mu$ equal to zero. The last means that in normal phase the density of sublattice A itinerant electrons is $n=1$. We have obtained four characteristic values of the applied field $H_{cr1}<H_3<H_0<H_4<H_{cr2}$. When $H_{cr1}<H<H_{cr2}$ the spin antiparallel p-wave superconductivity with $T_{1u}$ configuration coexists  with magnetism. The thermal superconductor to normal magnet transition is second order when $H$ runs the interval $(H_3,H_4)$. It is an abrupt transition when $H_{cr1}<H<H_3$ or $H_4<H<H_{cr2}$. The dimensionless gap $\Delta/t$ as a function of $H/H_0$ at zero temperature is depicted in Fig.(\ref{gap-h}) for parameters $J/t=4/0.3$ and $J^B/J=1/15$. The critical values of the external magnetic fields are $H_{cr1}/H_0=0.927$ and $H_{cr2}/H_0=1.062$. The red vertical lines in  Fig.(\ref{gap-h}) correspond to the $H_3/H_0=0.944$ and $H_4/H_0=1.045$.

\begin{figure}[!t]
\epsfxsize=\linewidth
\epsfbox{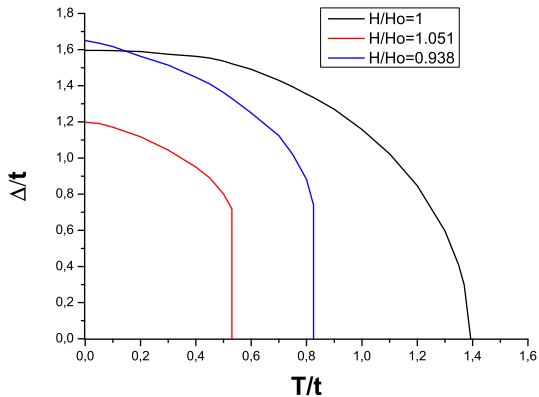} \caption{(Color online)\,\,Dimensionless gap $\Delta/t$ as a function of dimensionless temperature $T/t$. The black line represents the function for $H=H_0$, the blue line for $H=0.938H_0$ and the red line for $H=1.051H_0$. }\label{gap-T}
\end{figure}

To demonstrate the nature of the thermal superconductor-normal magnet transition, we have calculated the gap $\Delta/t$ as a function of the temperature $T/t$ for three different values of the external magnetic field: $H/H_0=0.89$, $H/H_0=1$ and $H/H_0=1.09$. The result is shown in Fig.(\ref{gap-T}).
The black line represents $\Delta/t$ as a function of $T/t$ for $H=H_0$. The second order phase transition demonstrates itself through the smooth decrease of the gap up to zero at critical temperature $T_{sc}=1.393t$. The other two lines, blue $H=0.938H_0$ and red $H=1.051H_0$, demonstrate abrupt fall of the gap at superconducting critical temperatures $T_{sc}=0.825\,t$ and  $T_{sc}=0.53\,t$ respectively.

To account for the Coulomb repulsion one has to add the Hubbard term to the Hamiltonian Eq.\eqref{IFerri1}
\be\label{IFerri13} h \rightarrow h+ U\sum\limits_{i\in A} c_{i\uparrow}^+c_{i\uparrow}c_{i\downarrow}^+c_{i\downarrow} \ee
When the coupling $ U/t$ is strong enough the model describes a system of localized electrons on sublattice A sites. Under the pressure, the charge screening increases (the Coulomb repulsion $U$ decreases) and overlap of the wave functions of electrons increases (the hopping parameter $t$ increases). As a result the coupling constant $U/t$ decreases and we can treat the Hubbard term in a weak coupling regime.

The contribution of the first order term in $U/t$ expansion to the magnon induced superconductivity changes the potential Eq.\eqref{IFerri7}
\be\label{IFerri14}
V_k\rightarrow V_k+U. \ee
It is known \cite{AK11} that this term do not contribute to any unconventional channel of superconductivity. To assess the suppression of the superconductivity due to  the first term we calculate the superconducting critical temperature as a function of $U/t$ for $H=H_0$, $J/t=10$ and $J^B/J=0.05$. When the Zeeman splitting is compensated ($H=H_0$) the thermal superconductor-normal magnet transition is a second order and we can use the linearized gap equation to determine the critical temperature $T_{sc}$ as a function of the Coulomb repulsion $U$
\be\label{IFerri15}
1 = \frac 13 \frac {1}{N^2} \sum\limits_{kp\in B_r}
\frac {\Gamma_k (V_{k-p}+U)\Gamma_p}{E_p} \tanh \frac {E_p}{2T_{sc}}.
 \ee
In equation \eqref{IFerri15} $\Gamma_k=\sin k_x+\sin k_y+\sin k_z$ and $E_p=2t|\cos p_x+\cos p_y+\cos p_z |$.

The result shows that the contribution of the first order term in $U/t$ expansion to the magnon induced superconductivity is unessential. For example for $U/t=0$ the critical temperature is $T_{sc}/t=1.393$, while for $U/t=0.8$ it is $T_{sc}/t=1.39$.

The higher order terms in a weak coupling expansion contribute to the superconductivity through the Kohn-Luttinger mechanism. The results show\cite{RKS10} that the effect on the p-wave superconductivity with $T_{1u}$ configuration is weak. This permits to conclude that the Coulomb repulsion, in a weak coupling regime, does not impact significantly the magnon induced superconductivity and we can drop it.

Finally, we consider the effect of the chemical manipulation. To this end we study the critical temperature $T_{sc}$ as a function of  the density of states of itinerant electrons in normal phase for the same parameters of  the system as above. The equation for the critical temperature $T_{sc}$ is the equation \eqref{IFerri15} with $U=0$ and $E_p=\sqrt{[2t(\cos p_x+\cos p_y+\cos p_z )+\mu]^2}$, where $\mu$ is the chemical potential.
The table shows that decreasing the density of itinerant electrons the superconducting critical temperature $T_{sc}/t$ slowly decreases. This is true if the electrons are delocalized. If the sublattice A electrons are localized the deviation from half-filling is a way to delocalize them and we expect an opposite tendency.
\begin{center}\begin{tabular}{|c|c|c|c|c|c|c|c|c|c|}\hline
$n$&1&0.9&0.8&0.7&0.6&0.5\\ \hline
$T_{sc}/t$&1.393&1.3873&1.3761&1.3534&1.3169&1.2823\\ \hline
\end{tabular}\end{center}
\vskip 0.25cm

In summary, we have proposed a method of preparation of superconducting ferrimagnetic spinel. We have  studied a two-sublattice spin-fermion model of ferrimagnetic spinel, with spin-$1/2$ itinerant electrons at the sublattice $A$ site and spin-$s$ localized electrons  at the sublattice $B$ site in an external magnetic field, applied along the magnetization of the localized electrons. Magnon induced superconductivity is predicted when the Coulomb repulsion is small (the system is under hydrostatic pressure) and the external magnetic field compensates the Zeeman splitting due to the spin-fermion exchange.

There are two methods of preparation of spinels. If, during the preparation, an external magnetic field as high as 300 O\"{e} is applied upon cooling the material is named field-cooled (FC). If the applied field is about 1O\"{e} the material is zero-field cooled (ZFC). The magnetization-temperature \cite{spinelFeCr2S4} and magnetic susceptibility \cite{spinel+,spinel++} curves for these materials display a pronounced bifurcation below N\'{e}el $T_N$ temperature. The (ZFC) curve exhibits a maximum and then a monotonic decrease upon cooling from $T_N$, while the (FC) curve increases steeply, shows a dip near the temperature at which the (ZFC) curve has a maximum and finally increases monotonically\cite{spinel++}. The magnetization-temperature curve is close to the reference curve obtained from contribution of localized spins on the one of the sublattices. Hence, in (FC) materials the electrons on the other sublattice have dispersion with approximately compensated Zeeman splitting. This permits to think that at high hydrostatic pressure these material will have a superconducting state.

The Zeeman splitting energy $H_0$ can be obtained from the external magnetic field used for the preparation of (FC) material. For $MnV_2O_4$ the field is as high as 300 O\"{e}\cite{spinel++}.

The model we have considered is a prototype model of itinerant ferrimagnetic spinel. It is prototype model because the sublattice A sites are occupied, usually, by more than one electron. But this is not a toy model, because it capture all physical relevant properties of the spinel system and the existence of more than one electrons on sublattice A sites will not change the conclusion that the magnon induced superconductivity exist upon some conditions.

Finally, we have not considered the $s$-band electrons because the spinel magnetism is determined by $d$-electrons. The goal of the paper is to study the  formation of Cooper pairs of delocalized  sublattice A d-electrons (the system is under hydrostatic pressure) in external magnetic field. The s-electrons are accounted for through the renormalization of the parameters of the spin-fermion model (\ref{IFerri1}).



\end{document}